\documentstyle[a4,12pt,epsfig]{article}
\begin{document}
\parskip=5pt plus 1pt minus 1pt

%\begin{flushright}
%\underline{\bf BIHEP-TH Preprint} 
%\end{flushright}

\vspace{0.2cm}

\begin{center}
{\Large\bf Four-Zero Texture of Hermitian Quark Mass Matrices and
Current Experimental Tests}
\end{center}

\begin{center}
{\bf Harald Fritzsch} 
\footnote{Electronic address: bm@theorie.physik.uni-muenchen.de} \\
{\it Sektion Physik, Universit$\it\ddot{a}$t M$\it\ddot{u}$nchen,
%Theresienstrasse 37A, 
D-80333 Munich, Germany}
\end{center}

\begin{center}
{\bf Zhi-zhong Xing} 
\footnote{Electronic address: xingzz@mail.ihep.ac.cn} \\
{\it CCAST (World Laboratory), P.O. Box 8730, Beijing 100080, China \\
and Institute of High Energy Physics, Chinese Academy of Sciences, \\
P.O. Box 918 (4), Beijing 100039, China}
\footnote{Mailing address}
\end{center}

\vspace{2cm}

\begin{abstract}
We show that the four-zero texture of Hermitian quark mass matrices
can survive current experimental tests: it is able to yield sufficiently 
large predictions for the flavor mixing parameter $|V_{ub}/V_{cb}|$ and 
the CP-violating parameter $\sin 2\beta$, without fine-tuning of the
input values of quark masses, provided the (2,2), (2,3) and (3,3) elements 
of up and down mass matrices roughly satisfy a geometrical hierarchy. 
The successful relations $|V_{us}| \approx \sqrt{m_u/m_c + m_d/m_s}$ and 
$|V_{td}/V_{ts}| \approx \sqrt{m_d/m_s}$ obtained previously from
the six-zero texture of Hermitian quark mass matrices remain unchanged
in the new ansatz. 
\end{abstract}

\newpage

\framebox{\Large\bf 1} ~
The structure of quark mass matrices, which governs the pattern of quark
flavor mixing, is completely unknown in the standard electroweak model.
A theory more fundamental than the standard model should be able to 
determine the quark mass matrices exclusively, such that the relevant
physical parameters (six quark masses, three flavor mixing angles and
one CP-violating phase) could be calculated. Attempts in this direction,
e.g., those starting from supersymmetric grand unification theories
and from superstring theories, are encouraging but have 
not proved to be very successful. Phenomenologically, a very common 
approach is to search for simple textures of quark mass matrices that can 
predict self-consistent and experimentally-favored relations between quark 
masses and flavor mixing parameters \cite{Review}. The discrete flavor 
symmetries hidden in such textures might finally provide useful hints 
towards the underlying dynamics responsible for quark mass generation 
and CP violation. 

An instructive ansatz of Hermitian quark mass matrices with six texture 
zeros was first proposed by one of us \cite{F78} about 25 years ago
%%%%%%%%%%%%%%%%%%%%%%%
\footnote{A pair of off-diagonal texture zeros of the up or down quark mass 
matrix have been counted, due to Hermiticity, as one zero \cite{RRR}.}:
%%%%%%%%%%%%%%%%%%%%%%%  
$$
M_{\rm u} \; = \; \left ( \matrix{
~ {\bf 0} & ~ C_{\rm u} ~	& {\bf 0} ~ \cr
~ C^*_{\rm u}	& ~ {\bf 0} ~	& B_{\rm u} ~ \cr
~ {\bf 0}	& ~ B^*_{\rm u} ~ & A_{\rm u} ~ \cr} \right ) \; ,
\eqno{\rm (1a)}
$$
$$
M_{\rm d} \; = \; \left ( \matrix{
~ {\bf 0} & ~ C_{\rm d} ~	& {\bf 0} ~ \cr
~ C^*_{\rm d}	& ~ {\bf 0} ~	& B_{\rm d} ~ \cr
~ {\bf 0}	& ~ B^*_{\rm d} ~ & A_{\rm d} ~ \cr} \right ) \; ,
\eqno{\rm (1b)}
$$
with $|A_{\rm q}| \gg |B_{\rm q}| \gg |C_{\rm q}|$ for q = u (up) or 
d (down). It predicts three simple relations between elements of the 
Cabibbo-Kobayashi-Maskawa (CKM) matrix and ratios of the quark masses:
$$
|V_{us}| \; \approx \; \left | \sqrt{\frac{m_u}{m_c}} - e^{i\phi_1} 
\sqrt{\frac{m_d}{m_s}} \right | \; , \;
\eqno{\rm (2)}
$$
and
$$
\frac{|V_{ub}|}{|V_{cb}|} \; \approx \; \sqrt{\frac{m_u}{m_c}} \;\; ,
~~~~~~~~~~~~~~~~
\eqno{\rm (3a)}
$$
$$
\frac{|V_{td}|}{|V_{ts}|} \; \approx \; \sqrt{\frac{m_d}{m_s}} \;\; ,
~~~~~~~~~~~~~~~~
\eqno{\rm (3b)}
$$
in which $\phi_1 \equiv \arg (C_{\rm u}) - \arg(C_{\rm d})$ denotes 
the phase difference between two (1,2) elements of $M_{\rm u}$ and 
$M_{\rm d}$. These relations keep to hold in a generalized version 
of $M_{\rm u}$ and $M_{\rm d}$ in Eq. (1), the so-called four-zero 
texture of Hermitian quark mass matrices \cite{FX95,4zero}:
$$
M_{\rm u} \; = \; \left ( \matrix{
~ {\bf 0} & ~ C_{\rm u} ~	& {\bf 0} ~ \cr
~ C^*_{\rm u}	& ~ \tilde{B}_{\rm u} ~	& B_{\rm u} ~ \cr
~ {\bf 0}	& ~ B^*_{\rm u} ~ & A_{\rm u} ~ \cr} \right ) \; ,
\eqno{\rm (4a)}
$$
$$
M_{\rm d} \; = \; \left ( \matrix{
~ {\bf 0} & ~ C_{\rm d} ~	& {\bf 0} ~ \cr
~ C^*_{\rm d}	& ~ \tilde{B}_{\rm d} ~	& B_{\rm d} ~ \cr
~ {\bf 0}	& ~ B^*_{\rm d} ~ & A_{\rm d} ~ \cr} \right ) \; ,
\eqno{\rm (4b)}
$$
where $|\tilde{B}_{\rm u}| \sim |B_{\rm u}| \sim m_c$ and
$|\tilde{B}_{\rm d}| \sim |B_{\rm d}| \sim m_s$ have commonly been
assumed. The key difference between two textures in Eqs. (1) and (4) is
in their predictions for $|V_{cb}|$; i.e.,
$|V_{cb}| \sim \sqrt{m_s/m_b}$ from the six-zero texture of 
$M_{\rm u}$ and $M_{\rm d}$ \cite{F78} and $|V_{cb}| \sim m_s/m_b$ from 
the four-zero texture of $M_{\rm u}$ and $M_{\rm d}$ \cite{FX95}. The 
former is apparently in conflict with the smallness of $|V_{cb}|$ 
extracted from current experimental data \cite{PDG}. That is why a lot 
of attention has been moved from six texture zeros to four
texture zeros of quark mass matrices \cite{Review}. 

However, the validity of Eq. (3a) has recently been questioned by several
authors (see, e.g., Refs. \cite{Kuo,Ross}). Their essential point is that
the magnitude of $|V_{ub}/V_{cb}|$ predicted by $\sqrt{m_u/m_c}$ is too 
low ($|V_{ub}/V_{cb}| \approx 0.06$ or smaller for reasonable values of 
$m_u$ and $m_c$ \cite{Leutwyler}) to agree with the present experimental 
result ($|V_{ub}/V_{cb}|_{\rm ex} \approx 0.09$ \cite{PDG}). To resolve this
discrepancy, two different approaches have been considered \cite{Ross}:
(a) revising the four-zero texture of $M_{\rm u}$ and $M_{\rm d}$ by 
introducing non-vanishing entries in their (1,3) and (3,1) positions, 
such that only the (1,1) elements vanish in the modified quark mass 
matrices \cite{Albright}; 
and (b) abandoning Hermiticity of the four-zero texture of $M_{\rm u}$
and $M_{\rm d}$, such that the (2,3) and (3,2) elements of either mass
matrix can be quite different in magnitude. Both approaches allow the
naive result in Eq. (3a) to be sufficiently enhanced, leading to good
agreement between quark mass matrices and current experimental data.

In this Letter we aim to point out that the four-zero texture of Hermitian
quark mass matrices can actually survive current experimental tests, if we 
give up the previous assumptions  
$|\tilde{B}_{\rm u}| \sim |B_{\rm u}| \sim m_c$ and
$|\tilde{B}_{\rm d}| \sim |B_{\rm d}| \sim m_s$, 
which are neither phenomenologically nor theoretically justified. 
We find that no fine-tuning of the input values of quark masses is needed 
to arrive at a favorable prediction for $|V_{ub}/V_{cb}|$, provided the 
(2,2), (2,3) and (3,3) elements of $M_{\rm u}$ and $M_{\rm d}$ in Eq. (4)
roughly satisfy a geometrical hierarchy.
The successful predictions obtained previously in Eq. (2) and Eq. (3b) 
remain unchanged, because they are insensitive to the hierarchy between 
$A_{\rm q}$ and its neighbors $B_{\rm q}$ and $\tilde{B}_{\rm q}$ 
(for q = u or d). As a result, the CP-violating parameter $\sin 2\beta$
also gets enhanced in our new ansatz. 

\vspace{0.3cm}

\framebox{\Large\bf 2} ~
Let us reconsider the four-zero texture of Hermitian quark mass 
matrices given in Eq. (4). The observed hierarchy of quark masses
($m_u \ll m_c \ll m_t$ and $m_d \ll m_s \ll m_b$) implies that
$|A_{\rm q}| \gg |\tilde{B}_{\rm q}|, |B_{\rm q}|, |C_{\rm q}|$ 
(for q = u or d) should in general hold \cite{F87}. It is well known 
that $M_{\rm q}$ can be decomposed into
$M_{\rm q} = P^\dagger_{\rm q} \overline{M}_{\rm q} P_{\rm q}$, where
$$
\overline{M}_{\rm q} \; = \; \left ( \matrix{
{\bf 0} & |C_{\rm q}|	& {\bf 0} \cr
|C_{\rm q}|	& \tilde{B}_{\rm q}	& |B_{\rm q}| \cr
{\bf 0}	& |B_{\rm q}|	& A_{\rm q} \cr} \right ) \; ,
\eqno{\rm (5)}
$$
and $P_{\rm q} = {\rm Diag} \left \{1, e^{i\phi^{~}_{C_{\rm q}}}, 
e^{i (\phi^{~}_{B_{\rm q}} + \phi^{~}_{C_{\rm q}})} \right \}$
with $\phi^{~}_{B_{\rm q}} \equiv \arg (B_{\rm q})$ and 
$\phi^{~}_{C_{\rm q}} \equiv \arg (C_{\rm q})$ is a pure phase matrix. 
For simplicity, we neglect the subscript ``q'' in the following,
whenever it is unnecessary to distinguish between the up and down quark 
sectors. The real symmetric mass matrix $\overline{M}$ can be diagonalized
by use of the orthogonal transformation
$$ 
O^{\rm T} \overline{M} O \; = \; \left ( \matrix{
\lambda_1 & 0 & 0 \cr
0 & \lambda_2 & 0 \cr
0 & 0 & \lambda_3 \cr} \right ) \; ,
\eqno{\rm (6)}
$$
where $\lambda_i$ (for $i=1,2,3$) are quark mass eigenvalues and may be
either positive or negative 
(i.e., $|\lambda_1| = m_1 \ll |\lambda_2| = m_2 \ll |\lambda_3| =m_3$). 
Without loss of generality, we take $\lambda_3 >0$ and $A >0$. 
Then ${\rm Det} (\overline{M}) = -A|C|^2 < 0$ implies that 
$\lambda_1 \lambda_2 < 0$ is required. Note that $\tilde{B}$, $|B|$
and $|C|$ can be expressed in terms of $\lambda_i$ and $A$ as follows:
$$
\tilde{B} \; = \; \lambda_1 + \lambda_2 + \lambda_3 - A \; , ~~~~~~
\eqno{\rm (7a)}
$$
$$ ~~~~~~~~~~~
|B| \; = \; \sqrt{\frac{(A -\lambda_1) (A -\lambda_2) 
(\lambda_3 -A)}{A}} \;\; ,
\eqno{\rm (7b)}
$$
$$
|C| \; = \; \sqrt{\frac{-\lambda_1 \lambda_2 \lambda_3}{A}} \;\; . 
~~~~~~~~~~~~
\eqno{\rm (7c)}
$$
The nine matrix elements of $O$ can also be calculated in terms of 
$\lambda_i$ and $A$. We obtain the exact analytical result as 
\small
$$
O \; =\; \left ( \matrix{ \cr
\displaystyle
\sqrt{\frac{\lambda_2 \lambda_3 (A-\lambda_1)}{A (\lambda_2 - \lambda_1)
(\lambda_3 - \lambda_1)}}
& \displaystyle
\eta  \sqrt{\frac{\lambda_1 \lambda_3 (\lambda_2 -A)}
{A (\lambda_2 - \lambda_1) (\lambda_3 - \lambda_2)}}
& \displaystyle
\sqrt{\frac{\lambda_1 \lambda_2 (A -\lambda_3)}
{A (\lambda_3 - \lambda_1) (\lambda_3 - \lambda_2)}} \cr\cr\cr
\displaystyle 
- \eta \sqrt{\frac{\lambda_1 (\lambda_1 -A)}{(\lambda_2 - \lambda_1)
(\lambda_3 - \lambda_1)}} 
& \displaystyle
\sqrt{\frac{\lambda_2 (A-\lambda_2)}
{(\lambda_2 - \lambda_1) (\lambda_3 - \lambda_2)}}
& \displaystyle
\sqrt{\frac{\lambda_3 (\lambda_3 -A)}
{(\lambda_3 - \lambda_1) (\lambda_3 - \lambda_2)}} \cr\cr\cr
\displaystyle
\eta \sqrt{\frac{\lambda_1 (A -\lambda_2) (A -\lambda_3)}
{A (\lambda_2 - \lambda_1) (\lambda_3 - \lambda_1)}}
& \displaystyle
- \sqrt{\frac{\lambda_2 (A-\lambda_1) (\lambda_3 -A)}
{A (\lambda_2 - \lambda_1) (\lambda_3 - \lambda_2)}}
& \displaystyle
\sqrt{\frac{\lambda_3 (A -\lambda_1) (A -\lambda_2)}
{A (\lambda_3 - \lambda_1) (\lambda_3 - \lambda_2)}} \cr\cr}
\right ) \; ,
\eqno{\rm (8)}
$$
\normalsize
where $\eta \equiv \lambda_2/m_2 = +1$ or $-1$, corresponding to
the possibility $(\lambda_1, \lambda_2) = (-m_1, +m_2)$ or
$(+m_1, -m_2)$. The CKM matrix, which measures the mismatch between
diagonalizations of $M_{\rm u}$ and $M_{\rm d}$, is given by
$V \equiv O^{\rm T}_{\rm u} (P_{\rm u} P^\dagger_{\rm d}) O_{\rm d}$.
Explicitly, we have 
$$
V_{i\alpha} \; =\; O^{\rm u}_{1i} O^{\rm d}_{1\alpha} +
O^{\rm u}_{2i} O^{\rm d}_{2\alpha} e^{i\phi_1} +
O^{\rm u}_{3i} O^{\rm d}_{3\alpha} e^{i(\phi_1 + \phi_2)} \; ,
\eqno{\rm (9)}
$$
where the subscripts $i$ and $\alpha$ run respectively over 
$(u,c,t)$ and $(d,s,b)$, and two phases are defined as
$\phi_1 \equiv \phi^{~}_{C_{\rm u}} - \phi^{~}_{C_{\rm d}}$ and
$\phi_2 \equiv \phi^{~}_{B_{\rm u}} - \phi^{~}_{B_{\rm d}}$.

There are two special but interesting patterns of $M_{\rm u}$ and 
$M_{\rm d}$, in which the matrix elements of $O^{\rm u}$ and 
$Q^{\rm d}$ depend purely upon the ratios of quark masses:
\begin{enumerate}
\item	If $\tilde{B}_{\rm u} = \tilde{B}_{\rm d} =0$ are taken, the 
four-zero texture of $M_{\rm u}$ and $M_{\rm d}$ is reduced to the 
six-zero texture \cite{F78}. In this case, $\eta =-1$ is uniquely fixed. 
Then $A_{\rm u}= m_u - m_c + m_t$ and $A_{\rm d}= m_d - m_s + m_b$ hold. 
The matrix elements of $O^{\rm u}$ and $Q^{\rm d}$ can be expressed,
respectively, in terms of $(m_u/m_c, m_c/m_t)$ and
$(m_d/m_s, m_s/m_b)$ \cite{Georgi}.
\item	Taking $\tilde{B}_{\rm u} = m_c$ and 
$\tilde{B}_{\rm d} = m_s$ \cite{22}, we are led uniquely to $\eta = +1$. 
In this case, $A_{\rm u} = m_t - m_u$ and $A_{\rm d} = m_b - m_d$ hold.
The matrix elements of $O^{\rm u}$ and $Q^{\rm d}$ can be expressed,
respectively, in terms of $(m_u/m_c, m_c/m_t)$ and
$(m_d/m_s, m_s/m_b)$ \cite{Review,22}.
\end{enumerate}
From both special patterns, however, one arrives finally at the 
questionable prediction given in Eq. (3a). We are therefore forced to 
start from Eq. (8), without any special assumption about the 
magnitudes of $\tilde{B}_{\rm u}$ and $\tilde{B}_{\rm d}$, to calculate 
the CKM matrix.

In view of the strong hierarchy of quark masses in both up and down 
sectors, we expect that $|\tilde{B}_{\rm u}| \ll A_{\rm u} \sim m_t$ 
and $|\tilde{B}_{\rm d}| \ll A_{\rm d} \sim m_b$ reasonably 
hold \cite{F87}. We emphasize that $|\tilde{B}_{\rm u}| \gg m_c$ and 
$|\tilde{B}_{\rm d}| \gg m_s$ are possible, a case which has not been 
considered before. Then we calculate the CKM matrix elements by use 
of Eqs. (8) and (9). To leading order, we obtain
%%%%%%%%%%%%%%%%%%%%%%%%%
\footnote{For simplicity, we have taken 
$\eta_{\rm u} = \eta_{\rm d} \equiv \eta = \pm 1$ in the analytical 
approximations. Our main results are essentially unchanged, even if 
$\eta_{\rm u} \neq \eta_{\rm d}$ 
(i.e., $\eta_{\rm u} = - \eta_{\rm d} = \pm 1$) is taken.}
%%%%%%%%%%%%%%%%%%%%%%%%
$$
|V_{ud}| \; \approx \; |V_{cs}| \; \approx \; |V_{tb}| 
\; \approx \; 1 \; , ~~~~~~~~~~~~~~~~~~
\eqno{\rm (10a)}
$$
$$
|V_{us}| \; \approx \; |V_{cd}| \; \approx \; 
\left | \sqrt{\frac{m_u}{m_c}} - e^{i\phi_1} \sqrt{\frac{m_d}{m_s}}
\right | \; , ~~~~~~
\eqno{\rm (10b)}
$$
$$ ~~~~~~~
|V_{cb}| \; \approx \; |V_{ts}| \; \approx \; 
\left | \sqrt{\frac{m_t - A_{\rm u}}{m_t}} - e^{i\phi_2} 
\sqrt{\frac{m_b - A_{\rm d}}{m_b}} \right | \; ;
\eqno{\rm (10c)}
$$
as well as
$$
\frac{|V_{ub}|}{|V_{cb}|} \; \approx \;
\left | \sqrt{\frac{m_u}{m_c}} - e^{i\phi} \frac{\eta \sqrt{m_d m_s}}
{m_b|V_{cb}|} \sqrt{\frac{m_b - A_{\rm d}}{m_b}} \right | \; ,
\eqno{\rm (11a)}
$$
$$
\frac{|V_{td}|}{|V_{ts}|} \; \approx \;
\sqrt{\frac{m_d}{m_s}} \;\; , ~~~~~~~~~~~~~~~~~~~~~~~~~~~~~~~~~~~
\eqno{\rm (11b)}
$$
where 
$$
\phi \; \equiv \; \arcsin \left ( \frac{\sin\phi_2}{|V_{cb}|}
\sqrt{\frac{m_t - A_{\rm u}}{m_t}} \right ) - \phi_1 \; . ~~~~~~
\eqno{\rm (12)}
$$
It is obvious that Eq. (10a), Eq. (10b) and Eq. (11b) are consistent 
with the previous results.  A goot fit of Eq. (10b) to the experimental
data suggests $\phi_1 \approx 90^\circ$ \cite{FX95,4zero}. Furthermore, Eq. (10c)
can trivially be arranged to agree with the present measurement,
because it involves three free parameter $A_{\rm u}$, $A_{\rm d}$
and $\phi_2$. However, it is nontrivial to simultaneously arrange
Eq. (10c) and Eq. (11a) to coincide with current data. The point
is that the factor $\sqrt{(m_b-A_{\rm d})/m_b}$ in Eq. (11a) should
be sufficiently large, such that $|V_{ub}/V_{cb}|$ can get enhanced and
reach its experimentally favored value. On the other hand, the term
$\sqrt{(m_t-A_{\rm u})/m_t}$ in Eq. (10c) must be sufficiently large too,
because the smallness of $|V_{cb}|$ requires significant cancellation
between the $\sqrt{(m_t-A_{\rm u})/m_t}$ term and the
$\sqrt{(m_b-A_{\rm d})/m_b}$ term. It turns out that 
$\phi_2 \approx 0^\circ$ is a good approximation, in order to
guarantee the sufficient cancellation between two big contributions
to $|V_{cb}|$. As a straightforward consequence, we have 
$\phi \approx -\phi_1 \approx -90^\circ$ from Eq. (12). Solving 
Eq. (11a), we find 
$$ ~~~~~~
A_{\rm d} \; \approx \; m_b \left [ 1 - |V_{cb}|^2 \frac{m_s}{m_d}
\left ( \frac{m_b}{m_s} \right )^2 \left (
\left | \frac{V_{ub}}{V_{cb}} \right |^2 - \frac{m_u}{m_c} \right ) 
\right ] \; .
\eqno{\rm (13)}
$$ 
Substituting Eq. (13) into Eq. (10c), we arrive at
$$ ~~~~~~~~~~~~~~~~
A_{\rm u} \; \approx \; m_t \left [ 1 - |V_{cb}|^2 \left (
\frac{m_b}{m_s} \sqrt{\frac{m_s}{m_d} \left (\left | 
\frac{V_{ub}}{V_{cb}} \right |^2 - \frac{m_u}{m_c} \right )}
~ \pm 1 \right )^2 \right ] \; .
\eqno{\rm (14)}
$$
Clearly there is no problem for Eq. (10c) and Eq. (11a) to agree with 
current experimental data, if $A_{\rm u}$ and $A_{\rm d}$ are given by 
Eqs. (13) and (14). Our results require no fine-tuning of the values of 
quark masses.

\vspace{0.3cm}

\framebox{\Large\bf 3} ~
To illustrate, let us take a more explicit and instructive example,
in which the previous prediction for 
$|V_{ub}/V_{cb}| \approx \sqrt{m_u/m_c}$ is enhanced by a factor 
of $\sqrt{2} ~$, i.e.,
$$
\frac{|V_{ub}|}{|V_{cb}|} \; \approx \;
\sqrt{2 \frac{m_u}{m_c}} \;\; . ~~~~~~~~~~
\eqno{\rm (15)}
$$
If $m_c/m_u \approx 265$ \cite{Leutwyler} is typically taken, one
gets $|V_{ub}/V_{cb}| \approx 0.087$ from Eq. (15), a result in good 
agreement with the experimental value 
$|V_{ub}/V_{cb}|_{\rm ex} = 0.089 \pm 0.008$ \cite{PDG}.
In this specific case, Eqs. (13) and (14) are simplified to
$$ 
A_{\rm d} \; \approx \; m_b \left [ 1 - |V_{cb}|^2 \frac{m_u m_s}
{m_c m_d} \left ( \frac{m_b}{m_s} \right )^2 \right ] \; . ~~~~~~~~~
\eqno{\rm (16a)}
$$
$$ 
A_{\rm u} \; \approx \; m_t \left [ 1 - |V_{cb}|^2 
\left ( \frac{m_b}{m_s} \sqrt{\frac{m_u m_s}{m_c m_d}} ~ 
\pm 1 \right )^2 \right ] \; . 
\eqno{\rm (16b)}
$$
Typically taking $m_c/m_u \approx 265$, 
$m_s/m_d \approx 20$ \cite{Leutwyler},
$m_b/m_s \approx 30$ \cite{Narison} and 
$|V_{cb}| \approx 0.04$ \cite{PDG}, we arrive at
$A_{\rm d}/m_b \approx 0.89$ and 
$A_{\rm u}/m_t \approx 0.86$ (corresponding to the plus sign) or
$0.92$ (corresponding to the minus sign). This numerical result 
is consistent very well with our original assumptions 
$A_{\rm u} \sim m_t$ and $A_{\rm d} \sim m_b$.
Taking account of Eq. (7a) and Eq. (7b), we get
$\tilde{B}_{\rm u}/A_{\rm u} \approx 0.16$ (plus sign) or
$0.09$ (minus sign) and $|B_{\rm u}|/A_{\rm u} \approx 0.4$ 
(plus sign) or $0.3$ (minus sign) as well as 
$\tilde{B}_{\rm d}/A_{\rm d} \approx 0.12$ and 
$|B_{\rm d}|/A_{\rm d} \approx 0.35$, where minor uncertainties
associated with $\eta =\pm 1$ in $\tilde{B}_{\rm u}$ and
$\tilde{B}_{\rm d}$ have been neglected. These results demonstrate
$$
|B_{\rm u}| \; \approx \; \sqrt{A_{\rm u} \tilde{B}_{\rm u}} 
\; \gg \; m_c \; ,
\eqno{\rm (17a)}
$$
$$
|B_{\rm d}| \; \approx \; \sqrt{A_{\rm d} \tilde{B}_{\rm d}} 
\; \gg \; m_s \; ,
\eqno{\rm (17b)}
$$
which are natural consequences of Eq. (7a) and Eq. (7b) 
in the leading-order approximation. Eq. (17) clearly indicates that
$A_{\rm q}$ and its neighbors $B_{\rm q}$ and $\tilde{B}_{\rm q}$
roughly satisfy a geometrical hierarchy:
$|B_{\rm q}|/A_{\rm q} \approx \tilde{B}_{\rm q}/|B_{\rm q}|$
(for q = u or d). 

Let us remark that the new prediction for $|V_{ub}/V_{cb}|$
in Eq. (15) results from the four-zero texture of Hermitian
quark mass matrices with a geometrical hierarchy among their 
(2,2), (2,3) and (3,3) elements, as illustrated in Eq. (17).
This new hierarchy has little influence on the other two
successful relations derived originally from the six-zero 
texture of Hermitian quark mass matrices, as shown in
Eq. (2) or Eq. (10b) and Eq. (3b) or Eq. (11b). Therefore
we expect that a value of $\sin 2\beta$ larger than before
can be predicted from our present ansatz, where 
$\beta \equiv \arg [- (V_{cd}V^*_{cb})/(V_{td}V^*_{tb})]$ is
an inner angle of the CKM unitarity triangle 
$$
V_{ud} V^*_{ub} + V_{cd} V^*_{cb} + V_{td} V^*_{tb} \; =\; 0
\eqno{\rm (18)}
$$
in the complex plane \cite{PDG}. We obtain
$$
\cos\beta \; \approx \; \frac{\displaystyle 
|V_{cd}|^2 + \frac{m_d}{m_s} - 2 \frac{m_u}{m_c}}
{\displaystyle 2 |V_{cd}| \sqrt{\frac{m_d}{m_s}}}
\eqno{\rm (19)}
$$
as a good approximation. We then arrive at 
$\beta \approx 22.5^\circ$ or $\sin 2\beta \approx 0.71$
using the typical inputs $|V_{cd}| \approx 0.221$ \cite{PDG},
$m_c/m_u \approx 265$ and $m_s/m_d \approx 20$ \cite{Leutwyler}.
This result is certainly compatible with the present 
measurements of CP violation in $B^0_d$ vs 
$\bar{B}^0_d \rightarrow J/\psi K_{\rm S}$ decays,
which yield $\sin 2\beta = 0.741 \pm 0.075$ (BaBar \cite{BaBar})
or $0.719 \pm 0.082$ (Belle \cite{Belle}). The other two
angles of the CKM unitarity triangle are found to be
$\alpha \equiv \arg [-(V_{td}V^*_{tb})/(V_{ud}V^*_{ub})]
\approx 77.1^\circ$ and 
$\gamma \equiv \arg [-(V_{ud}V^*_{ub})/(V_{cd}V^*_{cb})]
\approx 80.4^\circ$.
Of course, our numerical results are mainly for the purpose
of illustration. A careful analysis to higher order has to be
done, once the four-zero texture of Hermitian quark mass
matrices is confronted with more precise experimental data.

It is worth pointing out that the hierarchical structures of 
$\overline{M}_{\rm u}$ and $\overline{M}_{\rm d}$ will become 
more transparent, if they are expanded in terms of two new
parameters 
$\epsilon_{\rm u} \equiv |B_{\rm u}|/A_{\rm u}$ and
$\epsilon_{\rm d} \equiv |B_{\rm d}|/A_{\rm d}$:
$$
\overline{M}_{\rm u} \; \sim \; A_{\rm u} \left ( \matrix{
~ {\bf 0} & ~ \theta^3_{\rm u} ~	& {\bf 0} ~ \cr
~ \theta^3_{\rm u}	& ~ \epsilon^2_{\rm u} ~	
& \epsilon_{\rm u} ~ \cr
~ {\bf 0}	& ~ \epsilon_{\rm u} ~ 
& {\bf 1} ~ \cr} \right ) \; ,
\eqno{\rm (20a)}
$$
$$
\overline{M}_{\rm d} \; \sim \; A_{\rm d} \left ( \matrix{
~ {\bf 0} & ~ \theta^3_{\rm d} ~	& {\bf 0} ~ \cr
~ \theta^3_{\rm d}	& ~ \epsilon^2_{\rm d} ~	
& \epsilon_{\rm d} ~ \cr
~ {\bf 0}	& ~ \epsilon_{\rm d} ~ 
& {\bf 1} ~ \cr} \right ) \; ,
\eqno{\rm (20b)}
$$
where the small parameters 
$(\theta_{\rm u}, \theta_{\rm d})$ \cite{FX95} 
are related to $(\epsilon_{\rm u}, \epsilon_{\rm d})$ in 
the following way:
$$
\theta^2_{\rm u} \; \sim \; \epsilon^6_{\rm u} \; \sim \; 
\frac{m_u}{m_c} \; ,
\eqno{\rm (21a)}
$$
$$
\theta^2_{\rm d} \; \sim \; \epsilon^3_{\rm d} \; \sim \; 
\frac{m_d}{m_s} \; .
\eqno{\rm (21b)}
$$
Clearly $|V_{us}| \sim \sqrt{\theta^2_{\rm u} + \theta^2_{\rm d}}$ 
and $|V_{cb}| \sim |\epsilon_{\rm u} - \epsilon_{\rm d}|$ hold to 
leading order. While $|V_{td}/V_{ts}| \sim \theta_{\rm d}$ is simply 
expected, $|V_{ub}/V_{cb}| \sim \theta_{\rm u}$ gets modified due 
to the non-negligible contribution of 
${\cal O}(\theta^2_{\rm d}) \sim {\cal O}(\theta_{\rm u})$ 
from the down quark sector. We stress that $\epsilon_{\rm u}$ and 
$\epsilon_{\rm d}$ are not very small (indeed $\epsilon_{\rm u}
\approx 0.3$ or $0.4$ and $\epsilon_{\rm d} \approx 0.35$ in the
example taken above). One may speculate that the texture of 
$\overline{M}_{\rm u}$ and $\overline{M}_{\rm d}$ in Eq. (20)
could naturally be interpreted from a string-inspired model of
quark mass generation \cite{Ibanez} or from a new kind of flavor 
symmetry and its perturbative breaking \cite{Flavor}.

\vspace{0.3cm}

\framebox{\Large\bf 4} ~
In summary, we have shown that the four-zero texture of Hermitian 
quark mass matrices $M_{\rm u}$ and $M_{\rm d}$ can survive current 
experimental tests. It is actually easy to obtain sufficiently large 
predictions for the flavor mixing parameter $|V_{ub}/V_{cb}|$ and 
the CP-violating parameter $\sin 2\beta$, provided the (2,2), (2,3)
and (3,3) elements of $M_{\rm u}$ and $M_{\rm d}$ roughly satisfy
a geometrical hierarchy. On the other hand, our new ansatz can
reproduce the successful relations 
$|V_{us}| \approx \sqrt{m_u/m_c + m_d/m_s}$ and 
$|V_{td}/V_{ts}| \approx \sqrt{m_d/m_s}$ derived originally from
the six-zero texture of Hermitian quark mass matrices.

It is worth remarking that the phenomenological consequences 
of a specific texture of quark mass matrices depend not only upon 
the number of their vanishing entries but also upon the hierarchy 
of their non-vanishing entries. The former might not be preserved 
to all orders or at any energy scales in the unspecified interactions 
which generate quark masses and flavor mixing. Nevertheless, those
experimentally favored textures at low energy scales (such as the
one under discussion) are possible to provide enlightening hints
at the underlying physics of fermion mass generation and CP violation
at high energy scales.

\vspace{0.5cm}

One of us (Z.Z.X.) is grateful to IPPP in University of Durham,
where the paper was finalized, for its warm hospitality. This work
was supported in part by National Natural Science Foundation of China.

\vspace{1cm}

%\newpage

\end{document}